\documentclass[aps,prl,twocolumn,groupedaddress]{revtex4}
\usepackage{graphicx}

\begin{document}

\title{
Electronic structure of the $\sigma$-phase in paramagnetic Fe-V alloys.
Experimental and theoretical study.
}

\author{J. Cieslak}
\email[Corresponding author: ]{cieslak@novell.ftj.agh.edu.pl}
\author{J. Tobola}
\author{S. M. Dubiel}
\affiliation{Faculty of Physics and Applied Computer Science,
AGH University of Science and Technology, al. Mickiewicza 30, 30-059 Krakow, Poland}

\date{\today}

\begin{abstract}
The electronic structure of $\sigma$-phase Fe$_{100-x}$V$_x$ compounds with 33.3
$\le x \le 60.0$ was calculated from the charge self-consistent
Korringa-Kohn-Rostoker method.  For the first time, charge densities
$\rho_A(0)$ and electric field gradients were determined at Fe nuclei, that
occupy five nonequivalent lattice sites.  The highest $\rho_A(0)$ values were
found on sites A and D, and the lowest one on site B, the difference ranging
between 0.162 and 0.174 $\it s$-like electrons per Fe atom for $x = 33.3$ and
$x = 60$, respectively.  The calculated quantities combined with
experimentally determined site occupancies were successfully applied to analyze
$^{57}$Fe M\"ossbauer spectra recorded on a series of 8 samples in a
paramagnetic state.
\end{abstract}


\maketitle

Studies of complex intermetallic phases, where several nonequivalent
crystallographic sites exist, are of a great interest \cite{Sluiter95,Havrankova01,vibrat00,Joubert08}.  Especially
challenging in this context are disordered alloy systems with the complex
structure such as $\sigma$-phases.  The latter, whose archetypal example is
that in the Fe-Cr system \cite{Bergman54}, belongs to an important class of tetragonal
close-packed crystallographic structures \cite{Nelson89} viz.  the Frank-Kasper phases \cite{Frank59}.
They are characterized by a high coordination number (CN), varying from CN=12
to 15 in the $\sigma$-phase.  The unit cell (space group $P4_2/mnm$) contains
30 atoms distributed in a non-stoichiometric way over 5 crystallographic sites,
commonly labeled A, B, C, D and E.  The importance of this class of phases is
further reinforced by the fact, that they exhibit topological properties
similar to simple metallic glasses due to an icosahedral local arrangement \cite{Dzugutow97}.
Consequently, they can also be regarded as very good approximants for
dodecagonal quasicrystals \cite{Dzugutow97}.  Indeed, close similarity between vibrational
properties of a one-component $\sigma$-phase on one hand and a one-component
glass with the icosahedron local arrangement on the other, was recently
reported \cite{vibrat00}.

The complex structure and a lack of stoichiometry also cause the interpretation
of experimental results obtained for $\sigma$-phase samples to be a difficult
task.  Here, a $\sigma$-phase magnetism which can be found in some alloy
systems like Fe-Cr and Fe-V is of a particular interest \cite{Cieslak08a,Cieslak09}.  In order to
understand it properly, one should know sublattice magnetic properties.  An
access to the latter is experimentally possible with microscopic techniques
such as M\"ossbauer Spectroscopy (MS) or Nuclear Magnetic Resonance (NMR).  We
are not aware of using NMR in studying Fe-containing $\sigma$-phases, but MS
has been already applied and distinction between electronic properties of
nonequivalent sites was not obtained due to low resolution in the measured spectra
\cite{Cieslak08a,Cieslak09}.  However, backed by dedicated electronic structure calculations,
relevant information on Fe charge-densities and electric field gradients (EFG)
on all nonequivalent sites has been already provided in the case of the $\sigma$-Fe-Cr
compounds in a paramagnetic state \cite{Cieslak08b}.

In this Letter we report results obtained with similar calculations for the
$\sigma$-phase in Fe$_{100-x}$V$_x$ alloys, also in the paramagnetic state.  This
series of compounds is especially well-suited for verifying electronic structure
computations, since a composition range of the Fe-V $\sigma$-phase existence is
several times wider than that in the Fe-Cr system.  The non-spin polarized KKR
results have been used to analyze $^{57}$Fe-site M\"ossbauer spectra recorded
at 295 K (paramagnetic state) in the $\sigma$-phase compounds
Fe$_{100-x}$V$_x$ with 34.4 $\le x \le 59.0$.

\begin{figure*}[hbt]
\includegraphics[width=.99\textwidth]{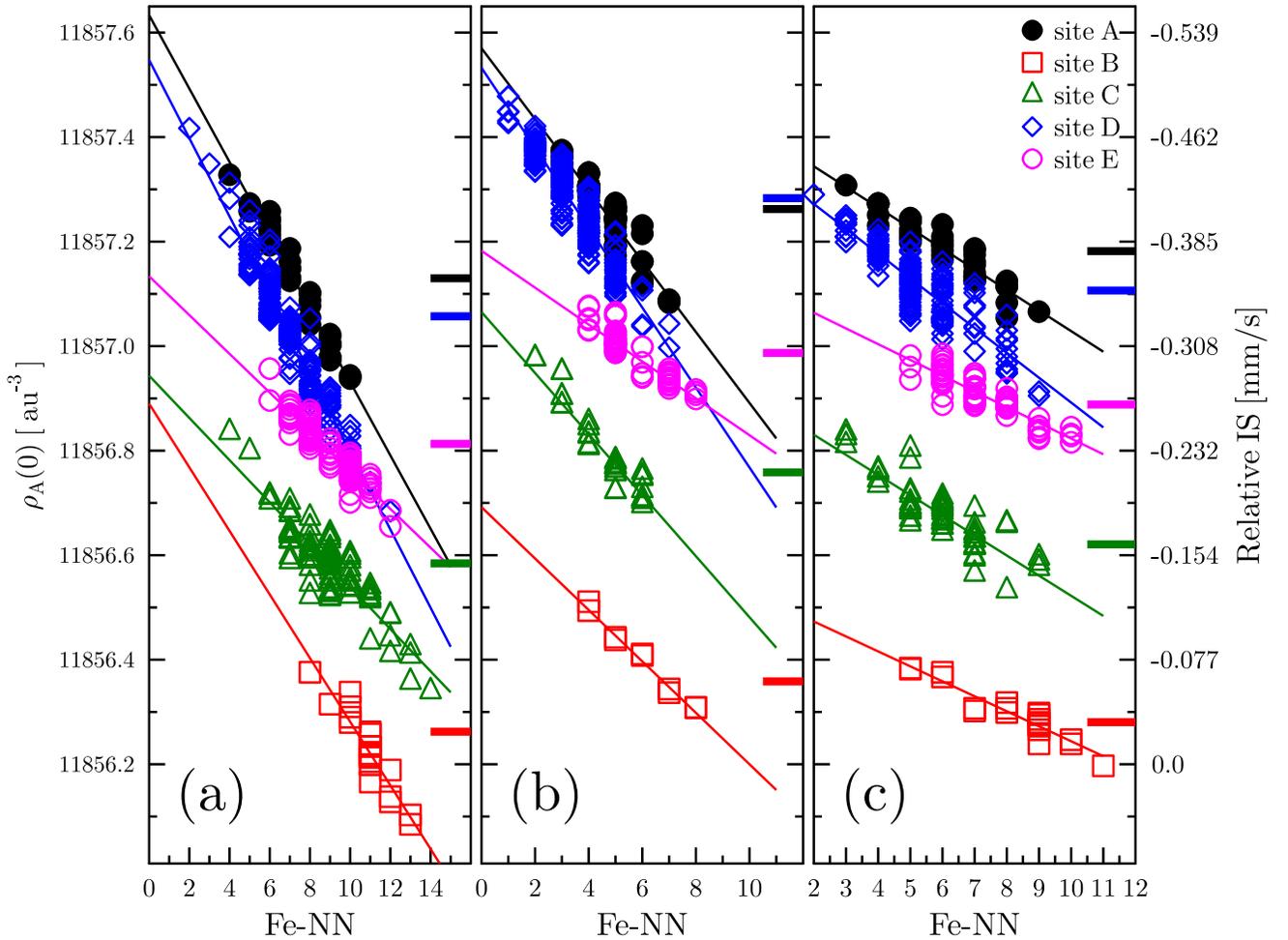}
\caption{(Online color)
Fe-site charge-density, $\rho_A(0)$, for five crystallographic sites versus the
number of NN-Fe atoms for (a) Fe$_{66.7}$V$_{33.3}$, (b) Fe$_{40}$V$_{60}$ and
(c) Fe$_{54}$Cr$_{46}$.  The lines stand for the best linear fits to the data
that represent $\rho_A(0)$-values for particular atomic NN-configurations.  The
average value of $\rho_A(0)$ for each site is indicated by a horizontal bar on
the right-hand axis on which are marked corresponding $IS$-values in mm/s.}
\label{fig1}
\end{figure*}

The M\"ossbauer spectrum associated with each site can be characterized by the
following spectral parameters:  amplitude $I$, line width $G$, isomer shift
(center of gravity), $IS$, that measures a charge-density at nucleus of the
probe atom, and quadrupole splitting, $QS$, that gives information on
the electric field gradient (EFG) in
the lattice.  Taking into account one parameter for a background and the fact
that in the unit cell of the sigma there are five sites (sublattices) one
needs 21 parameters to describe the overall spectrum.  However, as the
$^{57}$Fe-site M\"ossbauer spectrum of the $\sigma$-phase in the paramagnetic state
has no well-resolved structure \cite{Cieslak09}, unique determination of these parameters
from the spectrum analysis itself appears impossible and additional information
is required.  The latter can be partly obtained e.  g.  from a neutron
diffraction experiment that has not only confirmed the presence of the probe
$^{57}$Fe atoms on all 5 sites but also yielded precise knowledge on their relative
population on these sites \cite{Cieslak08c}.  Assuming the Lamb- M\"ossbauer factor to be
site independent, the relative contribution of each subspectrum ascribed to the
site should be equal to the corresponding values determined from the neutron
experiment.  Setting the latter as constraints, reduces the number of free
parameters from 21 to 17 which is still to high to allow the unique refinement
of the spectrum.  Theoretical calculations, as those presented elsewhere, are of a
great help in this respect \cite{Cieslak08b}.

It seems that the KKR method combined with the coherent potential approximation
(CPA) would be best adapted to investigate electronic structure of chemically
disordered $\sigma$-phase as it occurs in the Fe-V system.  However, such
self-consistent calculations are very time-consuming due to the presence of
Fe/V atoms disorder on all sites as well as a high multiplicity (8i) of three sites
giving rise to a long process of CPA self-consistency cycles.  In addition, such
results may only give us the quantities averaged over all possible atomic
configurations.  This can be important when studying theoretically magnetic
properties of these compounds, but it appears rather useless in the determination of
the spectral parameters in the paramagnetic state.
We have proposed an alternative approach to
face the problem \cite{Cieslak08b}, in which the electronic structure of the $\sigma$-phase was
calculated in a reasonably large number of ordered approximants, namely atomic
configurations.  These configurations should represent the most probable Fe and
V arrangements in the disordered $\sigma$-phase unit cell and cover wide range
of the Fe-nearest neighbour (NN) numbers but account for the fact that some
configurations are much less probable than others \cite{Cieslak08b}.  Consequently, we have
taken into accout those NN configurations whose overall probability was greater or equal
0.95.  Such probability cut-off has permitted to significantly reduce the
configurations number and diminish computation time to a reasonable level.  In
practice, all calculations were carried out using the lowest symmetry simple
tetragonal unit cell ($P1$), in which the atoms had the same positions as in
the original symmetry group ($P4_2/mnm$), but they were occupied either by Fe
or V atoms.  In the case of the Fe$_{100-x}$V$_x$ alloys, where the
$\sigma$-phase exists for $33 \le x \le 60$, two border compounds have been
chosen, i.e.  Fe$_{20}$V$_{10}$ corresponding to $x = 33.3$, and
Fe$_{12}$V$_{18}$ equivalent to $x = 60$.  In order to obtain the most probable
atomic configurations that satisfy the condition on the nearest neighborhood
appearing with the overall probability $\ge 0.95$, it was enough to take into
account 26 and 17 different configurations for the $x = 33.3$ and $x = 60$
compounds, respectively.  More details on the calculations can be found
elsewhere \cite{Cieslak08b}.

Fig.  1 presents the charge-density at $^{57}$Fe nuclei, $\rho_A(0)$, and the
corresponding isomer shift, $IS$, calculated for the specified $\sigma$-phase compositions for
nonequivalent sites.  Analyzing all dependences one can come to the following conclusions:
(i) $\rho_A(0)$ is characteristic of each site and it linearly decreases with
the Fe-NN number at a rate typical for each site and composition, (ii) the highest
charge-density have the sites A and D, and the lowest one the site B, and the
difference between them depends on the composition but hardly on the number of
Fe-NN atoms, (iii) the value of the charge-density at a given site depends on
the alloy composition.  The decrease of the charge-density with the increase of
the Fe-NN atoms observed for all sites and compositions means that there is an
effective charge transfer from V atoms to Fe atoms.  A similar effect was
observed for disordered $bcc$ Fe-V alloys \cite{Dubiel83}.  Using a fitted scaling
constant from \cite{Walker61}, one can correlate the calculated charge-densities with the
number of $s$-like electrons.  Thus the average charge density at Fe nuclei on
the site A is equal to 0.175 for $x = 33.3$ and to 0.205 $s$-like lectrons for $x = 60$.  For
site B the corresponding values are 0.013 and 0.031.  It is also interesting to
estimate in this way the difference in the charge-density between the extreme
sites i.e.  A and B.  By doing so one arrives at 0.162 for $x = 33.3$ and 0.174
for $x = 60$ i.e.  the difference hardly depends on the composition.

The charge-densities for any other $x$-value lying between 33.3 and 60 can be
determined as a weighted average of the parameters obtained for the two border
compositions.
Taking into account $IS$-values calculated for different compositions, atom
configurations and sites one can calculate an average $IS$- value, $<IS>_i$
($i$ = A, B, C, D, E) for each nonequivalent site.  The $<IS>_i$ values
obtained in this way are marked on the right-hand scale in Fig.  1 as
horizontal bars.  The probability of each site, $P_i$ as well as the
aforementioned $<IS>_i$ allow determining the average isomer shift for selected
composition $<IS>$,
\begin{equation}
<IS> = \sum P_i <IS>_i
\end{equation}
\begin{figure}[htb]
\includegraphics[width=.50\textwidth]{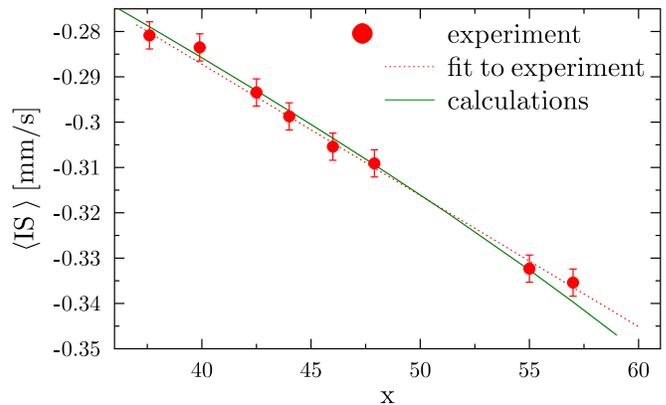}
\caption{(Online color)
Average center shift, $<CS>$ versus vanadium concentration, $x$, as measured
(circles) and as calculated (solid line).  The dashed line indicates the
best-fit to the measured values.}
\label{fig2}
\end{figure}
As illustrated in Fig.  2 theoretically calculated $<IS>$-values are in a good
agreement with the measured center shift, $<CS>$.  The data presented here also
give a clear evidence that substituting Fe by V in the $\sigma$-phase structure
has quite similar effect on the Fe-site charge-density values as already
observed in the $\alpha$-phase structure \cite{Dubiel83}, where it also increased the Fe-
site charge-density.

The good agreement between the experimental and calculated data give us a
confidence that our electronic structure calculations deliver reliable
description of the hyperfine parameters in the $\sigma$-phase samples.  One may
expect that a more detailed information obtained for the sublattice
charge-density is also credible.

\begin{figure}[htb]
\includegraphics[width=.50\textwidth]{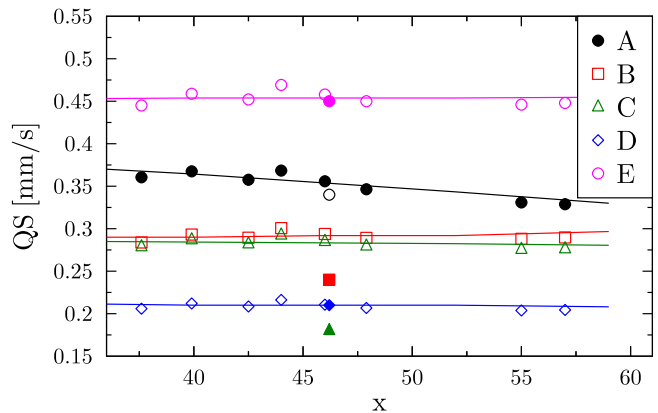}
\caption{(Online color)
Quadrupole splitting, $QS$, as determined for each site and vanadium
concentration, $x$, from the analysis of the measured spectra with the protocol
described in the text.  Values obtained previously for a $\sigma$-Fe-Cr
compound are indicated as reversely-filled symbols for comparison.}
\label{fig3}
\end{figure}
Concerning the second spectral parameter i.e.  $QS$, it was calculated based on
an extended point charge model as described in detail elsewhere \cite{Cieslak08b}.  The
obtained $QS$-values for each site and composition are shown in Fig.  3.  It is
clear from the figure that also $QS$ is characteristic of each site, maybe
except for the sites B and D for which $QS$-values are very close to each other whatever the
concentration $x$.  Another interesting feature is that except for the A site
the $QS$-values remain constant vs.  composition for all sites.  The greatest
$QS$-value was found for site E ($\sim$0.45 mm/s) and the lowest one for C
($\sim$0.20 mm/s).  For comparison, $QS$-values obtained previously for the
$\sigma$-Fe-Cr compound \cite{Cieslak08b} are marked by reversely-filled symbols in Fig.  3.
Interestingly, for the sites A, C and E they match very well the values
calculated in the $\sigma$-Fe-V, while for the sites B and D they are
significantly smaller and differ one from another.  Finally, using for the
relative intensities of the five subspectra the values as found from the
neutron diffraction experiment \cite{Cieslak08c} on one hand, and taking for $IS$ and $QS$ the
values obtained from the present calculations on the other hand, we have
successfully refined the measured $^{57}$Fe M\"ossbauer spectra.  In the
fitting procedure, each subspectrum was regarded as composed of double lines
with the same $QS$ and $IS$-values linearly dependent on the number of the
Fe-NN atoms and probabilities obtained from the binomial distribution.  It should be
mentioned that only five free parameters were needed for the analysis, i.e.
background, total spectral area, $IS$ for site B (to adjust the refined
spectrum to the used source of the gamma rays), line width and a
proportionality factor between $QS$ and an energy shift due to the quadrupole
interactions \cite{Cieslak08b}.  The agreement between the experimental and calculated
spectra is excellent for all measured compositions
of the $\sigma$-Fe-V system.

\begin{figure}[htb]
\includegraphics[width=.50\textwidth]{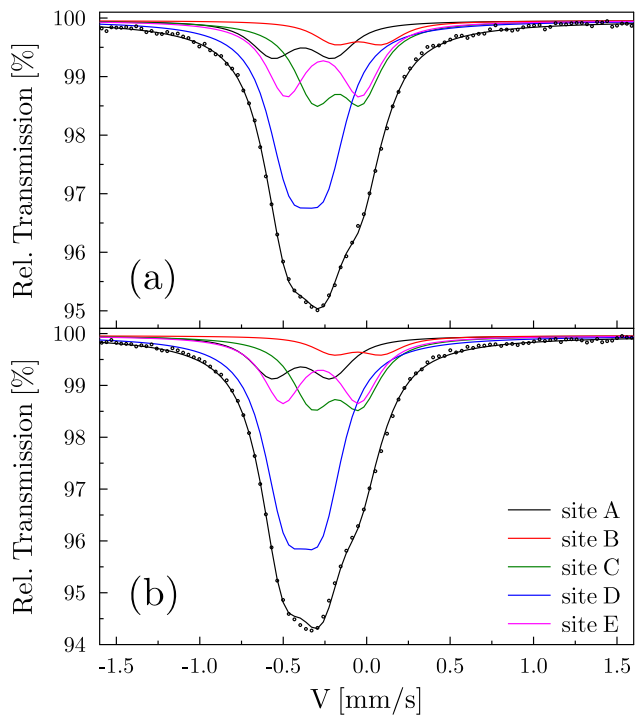}
\caption{(Online color)
$^{57}$Fe-site M\"ossbauer spectra recorded at 295 K on two samples with (a) $x
= 37$ and (b) $x = 46$.  The best-fit spectrum and five subspectra are
indicated by solid lines.}
\label{fig4}
\end{figure}

In summary, our results of the KKR electronic structure calculations obtained
for the $\sigma$-Fe-V system lead to conclusion that the proposed procedure,
based on the hyperfine parameters calculated from the first principles in combination with
the experimental data for the atom distribution over the five sublattices,
gives information on the charge-density and the electric field gradient in the complex disordered alloy system.
It should be mentioned here, that
this information was obtained for all five sublattices separatelly, taking into account the real
atomic concentrations on them.

The progress and adventage of our calculations in comparison to those carried
out by others, in which homogeneously occupied sublattices were considered,
is that we
have taken into account much more various atomic configurations that has enabled us to
get more deep, precise and realistic insight into the electronic structure of
the $\sigma$-phase. Our
calculations have been, in addition, verified by a comparison with experimental
data. A very good agreement achieved can be taken as evidence that the
procedure we have applied gives a good approach to the real electronic
structure of the
studied system. It can be also of a practical importance while analyzing
ill-resolved M\"ossbauer spectra recorded on samples with a complex structure.

\begin{acknowledgments}
The study was supported by The Ministry of Science and Higher Education,
Warszawa.
\end{acknowledgments}


\begin{thebibliography}{58}


\expandafter\ifx\csname natexlab\endcsname\relax\def\natexlab#1{#1}\fi
\expandafter\ifx\csname bibnamefont\endcsname\relax
  \def\bibnamefont#1{#1}\fi
\expandafter\ifx\csname bibfnamefont\endcsname\relax
  \def\bibfnamefont#1{#1}\fi
\expandafter\ifx\csname citenamefont\endcsname\relax
  \def\citenamefont#1{#1}\fi
\expandafter\ifx\csname url\endcsname\relax
  \def\url#1{\texttt{#1}}\fi
\expandafter\ifx\csname urlprefix\endcsname\relax\def\urlprefix{URL }\fi
\providecommand{\bibinfo}[2]{#2}
\providecommand{\eprint}[2][]{\url{#2}}

\bibitem{Sluiter95}
\bibinfo{author}{\bibfnamefont{M.}~\bibnamefont{Sluiter}},
  \bibinfo{author}{\bibfnamefont{K.}~\bibnamefont{Esfariani}},
  \bibnamefont{and} \bibinfo{author}{\bibfnamefont{Y.}~\bibnamefont{Kawazoe}},
  \bibinfo{journal}{Phys. Rev. Lett.} \textbf{\bibinfo{volume}{75}},
  \bibinfo{pages}{3142} (\bibinfo{year}{1995}).

\bibitem{Havrankova01}
\bibinfo{author}{\bibfnamefont{J.}~\bibnamefont{Havrankova}},
  \bibinfo{author}{\bibfnamefont{J.}~\bibnamefont{V\u{r}e\u{s}tal}},
  \bibinfo{author}{\bibfnamefont{L.~G.} \bibnamefont{Wang}}, \bibnamefont{and}
  \bibinfo{author}{\bibfnamefont{M.}~\bibnamefont{\u{S}ob}},
  \bibinfo{journal}{Phys. Rev. B} \textbf{\bibinfo{volume}{63}},
  \bibinfo{pages}{174104} (\bibinfo{year}{2001}).

\bibitem{vibrat00}
\bibinfo{author}{\bibfnamefont{S.~I.}~\bibnamefont{Simdyankin}},
  \bibinfo{author}{\bibfnamefont{S.~N.}~\bibnamefont{Taraskin}},
  \bibinfo{author}{\bibfnamefont{M.} \bibnamefont{Dzugutov}},
  \bibnamefont{and} \bibinfo{author}{\bibfnamefont{S.~R.}~\bibnamefont{Elliott}},
  \bibinfo{journal}{Phys.\ Rev.\ B} \textbf{\bibinfo{volume}{62}},
  \bibinfo{pages}{3223} (\bibinfo{year}{2000}).

\bibitem{Joubert08}
  \bibinfo{author}{\bibfnamefont{J.-M.}~\bibnamefont{Joubert}}
  \bibinfo{journal}{Progr. Mater. Sci.} \textbf{\bibinfo{volume}{53}},
  \bibinfo{pages}{528} (\bibinfo{year}{2008}).


\bibitem{Bergman54}
\bibinfo{author}{\bibfnamefont{G.}~\bibnamefont{Bergman}} \bibnamefont{and}
  \bibinfo{author}{\bibfnamefont{D.~P.} \bibnamefont{Shoemaker}},
  \bibinfo{journal}{Acta Cryst.} \textbf{\bibinfo{volume}{7}},
  \bibinfo{pages}{857} (\bibinfo{year}{1954}).

\bibitem{Nelson89}
  \bibinfo{author}{\bibfnamefont{D.~R.}~\bibnamefont{Nelson}}
  \bibnamefont{and}
  \bibinfo{author}{\bibfnamefont{F.}~\bibnamefont{Spaepen}},
  \bibinfo{journal}{Solid State Phys.} \textbf{\bibinfo{volume}{42}},
  \bibinfo{pages}{1} (\bibinfo{year}{1989}).

\bibitem{Frank59}
  \bibinfo{author}{\bibfnamefont{F.~C.}~\bibnamefont{Frank}}
  \bibnamefont{and}
  \bibinfo{author}{\bibfnamefont{J.~S.}~\bibnamefont{Kasper}},
  \bibinfo{journal}{Acta Crystalogr.} \textbf{\bibinfo{volume}{12}},
  \bibinfo{pages}{483} (\bibinfo{year}{1959}).

\bibitem{Dzugutow97}
  \bibinfo{author}{\bibfnamefont{M.}~\bibnamefont{Dzugutow}},
  \bibinfo{journal}{Phys. Rev. Lett.} \textbf{\bibinfo{volume}{79}},
  \bibinfo{pages}{4073} (\bibinfo{year}{1997}).

\bibitem{Cieslak08a}
  \bibinfo{author}{\bibfnamefont{J.}~\bibnamefont{Cieslak}},
  \bibinfo{author}{\bibfnamefont{M.}~\bibnamefont{Reissner}},
  \bibinfo{author}{\bibfnamefont{W.}~\bibnamefont{Steiner}}
  \bibnamefont{and}
  \bibinfo{author}{\bibfnamefont{S.~M.}~\bibnamefont{Dubiel}},
  \bibinfo{journal}{Phys. Stat. Sol.} \textbf{\bibinfo{volume}{205}},
  \bibinfo{pages}{1794} (\bibinfo{year}{2008}).

\bibitem{Cieslak09}
  \bibinfo{author}{\bibfnamefont{J.}~\bibnamefont{Cieslak}},
  \bibinfo{author}{\bibfnamefont{B.~F.~O.}~\bibnamefont{Costa}},
  \bibinfo{author}{\bibfnamefont{S.~M.}~\bibnamefont{Dubiel}},
  \bibinfo{author}{\bibfnamefont{M.}~\bibnamefont{Reissner}}
  \bibnamefont{and}
  \bibinfo{author}{\bibfnamefont{W.}~\bibnamefont{Steiner}},
  \bibinfo{journal}{J. Magn. Magn. Matter.} \textbf{\bibinfo{volume}{321}},
  \bibinfo{pages}{2160} (\bibinfo{year}{2009}).

\bibitem{Cieslak08b}
  \bibinfo{author}{\bibfnamefont{J.}~\bibnamefont{Cieslak}},
  \bibinfo{author}{\bibfnamefont{J.}~\bibnamefont{Tobola}},
  \bibinfo{author}{\bibfnamefont{S.~M.}~\bibnamefont{Dubiel}},
  \bibinfo{author}{\bibfnamefont{S.}~\bibnamefont{Kaprzyk}},
  \bibinfo{author}{\bibfnamefont{W.}~\bibnamefont{Steiner}}
  \bibnamefont{and}
  \bibinfo{author}{\bibfnamefont{M.}~\bibnamefont{Reissner}},
  \bibinfo{journal}{J. Phys.: Condens. Matter.} \textbf{\bibinfo{volume}{20}},
  \bibinfo{pages}{235234} (\bibinfo{year}{2008}).

\bibitem{Cieslak08c}
  \bibinfo{author}{\bibfnamefont{J.}~\bibnamefont{Cieslak}},
  \bibinfo{author}{\bibfnamefont{M.}~\bibnamefont{Reissner}},
  \bibinfo{author}{\bibfnamefont{S.~M.}~\bibnamefont{Dubiel}},
  \bibinfo{author}{\bibfnamefont{J.}~\bibnamefont{Wernisch}}
  \bibnamefont{and}
  \bibinfo{author}{\bibfnamefont{W.}~\bibnamefont{Steiner}},
  \bibinfo{journal}{J. Alloys Comp.} \textbf{\bibinfo{volume}{460}},
  \bibinfo{pages}{20} (\bibinfo{year}{2008}).

\bibitem{Dubiel83}
  \bibinfo{author}{\bibfnamefont{S.~M.}~\bibnamefont{Dubiel}}
  \bibnamefont{and}
  \bibinfo{author}{\bibfnamefont{W.}~\bibnamefont{Zinn}},
  \bibinfo{journal}{J. Magn. Magn. Matter.} \textbf{\bibinfo{volume}{37}},
  \bibinfo{pages}{237} (\bibinfo{year}{1983}).

\bibitem{Walker61}
  \bibinfo{author}{\bibfnamefont{L.~R.}~\bibnamefont{Walker}},
  \bibinfo{author}{\bibfnamefont{G.~K.}~\bibnamefont{Wertheim}}
  \bibnamefont{and}
  \bibinfo{author}{\bibfnamefont{V.}~\bibnamefont{Jaccarino}},
  \bibinfo{journal}{Phys. Rev. Letter} \textbf{\bibinfo{volume}{6}},
  \bibinfo{pages}{98} (\bibinfo{year}{1961}).

\bibitem{Gupta90}
  \bibinfo{author}{\bibfnamefont{A.}~\bibnamefont{Gupta}},
  \bibinfo{author}{\bibfnamefont{G.}~\bibnamefont{Principi}}
  \bibnamefont{and}
  \bibinfo{author}{\bibfnamefont{G.~M.}~\bibnamefont{Paolucci}},
  \bibinfo{journal}{Hyperfine Interact.} \textbf{\bibinfo{volume}{54}},
  \bibinfo{pages}{805} (\bibinfo{year}{1990}).

\end{thebibliography}
\end{document}